# Addressing Bias in Generative AI:
# Challenges and Research Opportunities in Information Management

Xiahua Wei
xhwei@uw.edu
School of Business
University of Washington, Bothell

Naveen Kumar
naveen.kumar@ou.edu
Price College of Business
University of Oklahoma, Norman

Han Zhang
han.zhang@scheller.gatech.edu
Scheller College of Business
Georgia Institute of Technology

# Addressing Bias in Generative AI:
# Challenges and Research Opportunities in Information Management

## Abstract

Generative AI technologies, particularly Large Language Models (LLMs), have transformed information management systems but introduced substantial biases that can compromise their effectiveness in informing business decision-making. This challenge presents information management scholars with a unique opportunity to advance the field by identifying and addressing these biases across extensive applications of LLMs. Building on the discussion on bias sources and current methods for detecting and mitigating bias, this paper seeks to identify gaps and opportunities for future research. By incorporating ethical considerations, policy implications, and sociotechnical perspectives, we focus on developing a framework that covers major stakeholders of Generative AI systems, proposing key research questions, and inspiring discussion. Our goal is to provide actionable pathways for researchers to address bias in LLM applications, thereby advancing research in information management that ultimately informs business practices. Our forward-looking framework and research agenda advocate interdisciplinary approaches, innovative methods, dynamic perspectives, and rigorous evaluation to ensure fairness and transparency in Generative AI-driven information systems. We expect this study to serve as a call to action for information management scholars to tackle this critical issue, guiding the improvement of fairness and effectiveness in LLM-based systems for business practice.

## 1. Introduction

Generative AI (GenAI), particularly Large Language Models (LLMs), is a transformative technology with unprecedented capabilities in natural language processing (NLP), content generation, and a myriad of other applications (Abbasi et al., 2024; IBM, 2024; Chowdhery et al., 2023; Touvron et al., 2023). Leveraging vast multimodal datasets that encompass text, images, audio, and video, GenAI produces new content based on patterns learned from training data (Yin et al., 2024). Prominent examples include GPT-4 for text generation and DALL-E and Midjourney for image creation. GenAI has sparked significant enthusiasm recently due to its potential to revolutionize business operations, drive efficiency, and create value across industries (Subramanian, 2024).

Despite its promising advantages, implementing GenAI is fraught with significant challenges, particularly concerning biases inherent in its data and algorithms (Chamberlain, 2024). Unlike traditional classification systems, which primarily categorize data into predefined classes, GenAI's generative nature amplifies the risks of perpetuating biases in training data and algorithms, thereby reinforcing gender, racial, and cultural stereotypes. These biases undermine trust and pose ethical, reputational, and regulatory risks for businesses and society.

As LLMs play a bigger role in business decision-making across functions such as operations, finance, marketing, and human resources (e.g., Davis et al., 2024; Manis and Madhavaram, 2023), they must align with human preferences (Shankar et al., 2024). Misalignment can lead to biased outcomes and unfair decisions (Konsynski et al., 2024; Dai et al., 2024). For instance, biased recruitment algorithms might favor specific genders (An et al., 2024), and biased models in healthcare could exacerbate inequities in patient care (Haltaufderheide and Ranisch, 2024).

As a result, bias in GenAI has become a pressing concern (Abbasi et al., 2024). A recent survey revealed that 32% of respondents believe they lost opportunities, such as financial or job prospects, due to biased AI algorithms. Additionally, 40% feel that companies using GenAI are not sufficiently protecting consumers from bias and misinformation (TELUS, 2023). Companies also face challenges in leveraging GenAI effectively, with model bias and trust being primary obstacles to its successful implementation (Deloitte, 2024). Given the crucial role of GenAI systems in information management, it is paramount for researchers and practitioners to understand and mitigate the biases these systems may perpetuate, ensuring they remain fair, equitable, and trustworthy.

Bias in GenAI manifests in multiple forms, including gender, racial, cultural, and ideological biases (Zhao et al., 2023). These biases frequently arise from both non-human and human factors embedded in training data and algorithms, reflecting societal prejudices and inequities (Hutchinson and Mitchell, 2019). Each type of bias is not unique to a specific business function; rather, they often intertwine across business practices, adding complexity to their detection and mitigation. This complexity requires explainable AI, a set of processes and methods that enable transparency and trust in the outcome of complex learning algorithms, to enable stakeholders to better understand how decisions are made and how biases can be identified. Additionally, policy frameworks often do not keep pace with the rapid advancements in GenAI technology, requiring proactive measures for effective regulation. Although some research has highlighted these concerns (Ray et al., 2024; Omrani et al., 2023; Mei et al., 2023), there remains a noticeable gap in studies systematically addressing bias in GenAI.

This paper explores the critical issue of bias in GenAI (LLMs), especially its implications for information management research. By examining current research and real-world examples, we first seek to understand the origins, measures, and impacts of LLM bias, as well as the techniques and challenges in mitigating it (Section 2). Our primary focus is on proposing directions for future information management research, including implementation strategies (Section 3) and business applications (Section 4), with each suggesting specific research questions for future studies. While addressing LLM bias may initially seem like a purely technical issue, it is profoundly intertwined with our evolving social and cultural environments. We advocate for interdisciplinary frameworks and methodologies with a dynamic lens to identify bias, enhance fairness, and reduce social harm. Moreover, we recommend incorporating proposed frameworks and methodologies into the research on various business practices within information management.

Our conceptual framework is summarized in Figure 1. Through these concerted efforts, our goal is to synthesize existing research on LLM bias, offer an innovative view to guide information management research in this vital and challenging area and guide the creation of LLM systems that uphold ethical integrity and inclusivity. While our review of technical methods, including bias measures and debiasing techniques, centers on text-based LLMs—the most prevalent format—our proposed framework for future research is generalizable to multimodal LLMs, including those using images, audio, and video.

## 2. Background and Context
In this section, we begin by defining bias in GenAI (LLM) models and explain how systematic errors can perpetuate favoritism and stereotypes. This is followed by an analysis of bias sources, encompassing both data-related and algorithmic factors. We then delve into methods for detecting and quantifying bias and summarize debiasing techniques aimed at mitigating these biases.

### 2.1. Defining Generative AI Bias
Bias in GenAI models refers to systematic errors or distortions in the model's processing of information, leading to favoritism towards certain groups or incorrect assumptions based on learned patterns (Mehrabi et al., 2021). These models, trained on vast corpora from diverse sources, inherit and amplify the biases in the data and algorithms (Bommasani et al., 2023). For instance, models like GPT-4o and their predecessors may generate outputs that prioritize certain perspectives, reinforce societal prejudices, and allocate opportunities unfairly if the training data predominantly reflects specific demographics or viewpoints (Shahriar et al., 2024).

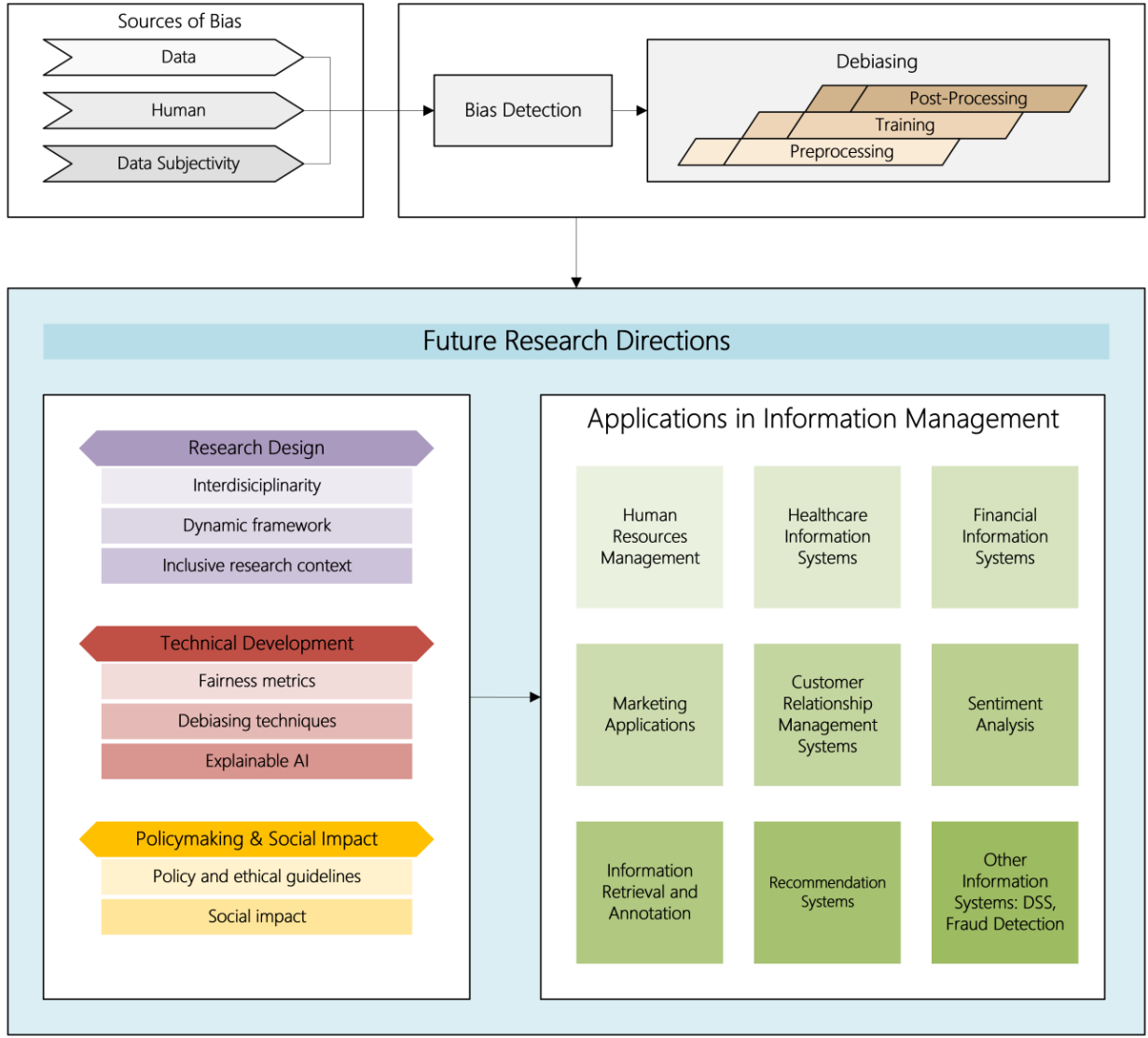

**Figure 1: Conceptual Framework**

### 2.2. Sources of Bias

Bias in LLMs is rooted in the foundations of artificial intelligence, arising from multiple interconnected sources, including data, algorithms, and human subjectivity (Susarla et al., 2023). While some biases stem from non-human factors, others are shaped by human decisions. Recognizing these origins, as well as their distinctions and relationships, allows information management scholars to comprehend the multifaced nature of bias and develop more targeted strategies for identifying and mitigating biases throughout the LLM lifecycle.

First, training *datasets* often contain inherent biases due to the source material. When data predominantly represents certain demographics or perspectives, models trained on them inevitably reflect and perpetuate these biases (Bommasani et al., 2023). Second, *algorithms* themselves can amplify bias through their inherent properties, such as mathematical assumptions, statistical properties, or the autonomous learning behaviors of complex models. These factors may introduce biases even when the training data seems unbiased.

*Human* subjectivity, meanwhile, exerts a pervasive influence that often exacerbates biases across both stages of LLM development (data and algorithm). During the data preparation phase, human-driven processes like labeling and annotation can skew representations that favor (or disfavor) certain demographics or scenarios (Gautam and Srinath, 2024). Similarly, when developing algorithms for product design, developers may prioritize certain user groups while neglecting others (Shulman and Gu, 2024). Additionally, algorithms may disproportionally emphasize patterns prevalent in majority groups, marginalizing minorities (Hovy and Prabhumoye, 2021). For instance, a language model designed primarily for Western audiences might perform poorly for users from other cultural backgrounds.

## 2.3 Detecting and Quantifying Bias

Detecting and quantifying bias in LLMs, as the first step to reducing bias, is a challenging yet critical area of research in information management. Various methodologies have been established in this area, many of which build on each other and evolve together. This interconnectedness reflects the complexity of bias detection and the need for a multifaceted approach to effectively address it in future research.

*Embedding-based metrics* measure bias by calculating conceptual distances between target words (e.g., nationalities) and attributes (e.g., races). While the Word Embedding Association Test uses cosine similarity to assess bias between word pairs (Caliskan et al., 2017), the Sentence Encoder Association Test extends this to sentence embeddings, capturing bias in more complex linguistic structures (e.g., Dolci et al., 2023). In addition, *probability-based* approaches measure systematic deviations from an unbiased outcome using probabilistic models and statistical inference such as Bayesian networks. These methods have been used to quantify racial, gender, and other discriminations in hiring, pay gaps, and criminal justice (Mehrabi et al., 2021).

Furthermore, the *counterfactual* evaluation tests for bias by modifying sentences, such as changing demographic indicators or key attributes, and then observing changes in model outputs. Significant bias impacts on gender and race have been identified using these methods (Boyer et al., 2023; Kusner et al., 2017). Additionally, *template-based* approaches evaluate biases by using predefined templates that vary demographic attributes while holding other variables constant, isolating their effect on outputs (Dong et al., 2024). Extensive templates have been created to assess various biases (Stanczak and Augenstein, 2021), with recent advancements in models to automatically generate prompts, reducing manual effort (Radcliffe et al., 2024).

## 2.4. Debiasing

Debiasing aims to enhance the accuracy of model predictions and recommendations while ensuring equity across different demographic groups (Susarla et al., 2023; Subramanian et al., 2021). Debiasing techniques can be applied at various stages of LLM implementation: the Preprocessing Stage, the Training Stage, and the Post-processing Stage. Below we discuss the existing debiasing techniques in each stage that have been established in the literature.

Data used to train or finetune LLMs can introduce downstream bias, making debiasing at the *preprocessing stage* critical, especially with imbalanced datasets. One common technique is Counterfactual Data Augmentation (CDA), which rebalances data by altering specific associations (Lu et al., 2020). For instance, sentences like "She is a nurse" can be replaced with "He is a nurse" to promote gender neutrality. Similarly, Counterfactual Data Substitution involves probabilistically replacing biased terms without altering the dataset size, mitigating bias while preserving the data structure (Maudslay et al., 2019). Other approaches remove biased examples from datasets, though this risks data loss and reduced coverage (Le Bras et al., 2020). An alternative method involves masking biased model weights during testing, allowing the model to bypass bias without the need for finetuning (Du et al., 2021). Further, training diverse datasets—such as across multiple languages—helps reduce ethnic and cultural biases, offering broader perspectives in model predictions (BehnamGhader and Milios, 2022). These techniques collectively can mitigate bias at the preprocessing stage, ensuring more balanced and fairer LLM outputs.

Debiasing during the *training stage* often involves regularization terms and various loss functions. One such method, contrastive loss, enables the model to distinguish between similar and dissimilar examples, thus minimizing reliance on biased features and promoting more balanced representations for long-term bias mitigation (He et al., 2022). Regularization techniques, such as dropout, focus on essential information and prevent models from learning irrelevant associations by introducing perturbations (Chen et al., 2024; Sokolová et al., 2024). In addition, contrastive learning enhances the model's ability to differentiate between biased and unbiased patterns (Oh et al., 2022; Li et al., 2023). Adversarial training introduces adversarial examples during training to challenge and reduce biased predictions (Zhang et al., 2018). Moreover, data auditing is crucial in bias detection, utilizing techniques like Datasheets for Datasets (Gebru et al., 2021) to document dataset origins and potential biases, ensuring transparency throughout the training process. Recent advancements also include auxiliary models that predict biased samples and perform sample reweighting during training, ensuring biased examples receive appropriate consideration (Chinta et al., 2024).

*Post-processing* debiasing techniques aim to identify and mitigate biases before GenAI systems are deployed. Bias auditing rigorously tests AI systems for biases, ensuring potential biases are addressed early to reduce the risk of biased outcomes in real-world applications (Raji et al., 2020). Additionally, human-LLM collaborations enhance the fairness and reliability of LLMs by integrating human input, feedback, and oversight throughout development and deployment. These collaborations allow for real-time monitoring and adjustment, with human reviewers providing iterative feedback to balance performance with bias mitigation (Ferdaus et al., 2024). This approach is particularly crucial in high-stakes applications, such as healthcare and law (Ma et al., 2024). Another key strategy is prompt engineering, which enhances debiasing efficiency by controlling how models generate responses (Schick et al., 2021). Techniques like zero-shot prompting, few-shot prompting, and chain-of-thought prompting have been shown to reduce bias and improve reasoning capabilities, leading to more equitable outcomes in GenAI systems (Kaneko et al., 2024).

### 3. Future Directions for Information Management Research
Addressing bias in LLMs is both an ethical and technical imperative, especially as these models play an increasingly prominent role in information systems. Firms may prioritize high-risk areas or those with regulatory requirements for bias mitigation, trading off other areas and leaving them under-addressed due to resource constraints. As a result, the challenges posed by LLM bias, such as perpetuating inequalities and unfair practices, demand robust mitigation strategies for bias and practical applications. Information management scholars are uniquely positioned to lead this effort. Future research should not only advance technical methods for reducing LLM bias but also explore how organizational and individual efforts can address these biases within real-world constraints. Developing frameworks that tackle the complex, multifaceted nature of LLM bias will be essential. Such efforts can enhance fairness, transparency, and equity, which ultimately improve individual, organizational, and societal benefits of LLMs, demonstrating their positive contribution to information management.

Strategies for advancing research in LLM bias encompass three interconnected areas: *research design*, *technical development*, and *policymaking/social impact*. Improved research design focuses on frameworks that capture the nuances of bias across various contexts, incorporating diverse perspectives (Section 3.1). Methodological advancements involve creating and refining metrics and algorithms to detect and mitigate LLM bias, ensuring greater accuracy and fairness (Section 3.2). Policymaking efforts aim to establish comprehensive guidelines and regulations that promote ethical practices in deploying LLMs, ensuring accountability and transparency (Section 3.3).

These three strategies are intricately related. Technical methodologies—such as bias quantification and debiasing techniques—are typically rooted in scientific research, but require an understanding of research design and policymaking/social impact. This interdependence arises because bias exists in a social context. Technical development is meaningful only within the context where the bias occurs. Whether quantifying bias or developing debiasing techniques, efforts need to be grounded in the social context to be effective.

By integrating these strategies, information management researchers can tackle the multifaceted challenges of LLM bias. Stakeholders within the LLM ecosystem—such as developers, companies, users, and policymakers—often have diverse and sometimes conflicting priorities. Recognizing and balancing these competing objectives is necessary to create more equitable LLMs.

### 3.1 Research Design

*Interdisciplinary Approaches and Collaboration:* Addressing LLM bias spans multiple domains, including computer science and engineering, information systems and information management, ethics, law, psychology, behavioral economics, sociology, political science, and other social sciences (Jiao et al., 2024). This interdisciplinary approach integrates diverse perspectives from stakeholders with distinct priorities. For instance, firms prioritize productivity and cost-effectiveness, end-users and advocacy groups often advocate for equity and inclusion, while policymakers emphasize fairness, transparency, and ethics. Computer scientists optimize performance, social scientists explore systemic inequalities perpetuated by LLMs, and legal experts assess liability and accountability. This challenge presents a unique opportunity for information management scholars to embrace an interdisciplinary approach. By collaborating with experts from diverse fields, they can create holistic solutions to tackle the complex nature of LLM bias, enhancing the fairness and accountability of information systems (Narayan et al., 2024).

The following research questions explore how interdisciplinary collaboration can help address the technical and societal dimensions of LLM bias: (1) What frameworks can information management scholars develop to reconcile tensions between technical priorities (e.g., performance and efficiency) and societal imperatives (e.g., fairness and transparency) in ways that address LLM bias unique in specific research contexts? (2) How can interdisciplinary research studies be designed to evaluate, prioritize, and trade off competing stakeholder objectives to create effective and equitable debiasing strategies? By answering these questions, we can develop collaborative approaches to detect and mitigate biases in LLMs while promoting fairness across diverse contexts.

*Dynamic Framework and Continuous Monitoring:* Bias mitigation in LLMs is not a one-time task but an ongoing process requiring continuous monitoring and adaptation as social norms and languages evolve (Brown, 2024). Practical constraints, such as high computational costs (especially in cloud computing), lack of expertise, and time pressures for extensive model retraining or continuous bias monitoring—which may conflict with regulatory requirements—often hinder effective resource-intensive debiasing efforts, especially for smaller organizations that often rely on generic solutions and prioritize rapid deployment over fairness. Additionally, constant monitoring demands continuous resources and iterative updates, which can potentially compromise extensive bias testing, complicating sustained bias mitigation efforts. To address these challenges, information management scholars must develop adaptive, resource-efficient frameworks that account for these constraints while maintaining effectiveness. For instance, lightweight monitoring tools and scalable debiasing techniques can alleviate computational burdens.

The following research questions focus on developing dynamic frameworks and continuous monitoring mechanisms under practical constraints to address the evolving nature of LLM bias: (1) What dynamic methodologies can information management scholars design to integrate real-time user feedback for continuous detection and mitigation of LLM bias, ensuring scalability and operational efficiency? (2) How can scholars develop and implement resource-efficient longitudinal studies to evaluate the effectiveness of continuous monitoring frameworks in mitigating bias over time while addressing expertise gaps and computational limitations? Addressing these questions will facilitate the creation of robust frameworks for dynamic feedback and longitudinal studies, ensuring LLM systems remain fair, transparent, and responsive to emerging biases over time.

*Inclusive Design and Cross-Cultural Studies:* Addressing LLM bias requires an understanding of how biases manifest across diverse communities and cultural contexts (Liu, 2024). However, complexity arises in determining whose perspectives take precedence. For instance, stakeholders from different cultural or socioeconomic backgrounds often possess conflicting definitions of fairness, making it challenging to establish universally acceptable solutions. As a result, a fairness criterion suitable in one context may inadvertently marginalize others. To navigate these challenges, information management researchers should explore methods that incorporate diverse stakeholder perspectives into the LLM design process. Engaging stakeholders from underrepresented and marginalized groups and across cultures ensures that LLM systems reflect a wide range of lived experiences and social contexts.

The following research questions explore how inclusive design and cross-cultural perspectives can mitigate LLM bias and foster equity in global applications: (1) How can the needs and perspectives of underrepresented and marginalized groups be effectively captured in the research design and data collection processes while balancing inclusivity, scalability, and the operational efficiency required for global deployment? (2) How does LLM bias manifest across national and cultural contexts, and what design strategies, grounded in information management theories, can balance the tension between culturally specific adaptations and the need for consistent global standards? Scholars can foster fairness, accountability, and positive societal impacts across diverse environments by exploring cultural dimensions of bias and crafting context-sensitive strategies.

### 3.2 Technical Development
*Improved Fairness Metrics:* While existing metrics provide a foundation for identifying LLM bias, they require further refinement to capture its complex and nuanced nature. Key challenges include balancing generalizability with context-specificity, adapting to evolving societal norms and biases, and managing computational costs associated with running comprehensive tests across multi-dimensional fairness evaluations. Information management scholars can develop adaptive, interpretable, and resource-efficient fairness metrics that account for diverse user groups and contexts while remaining practical for real-world implementation.

We propose the following research questions to advance this area: (1) How can fairness metrics balance generality and specificity to ensure broad applicability while capturing nuanced and context-specific biases? (2) What resource-efficient methodologies can operationalize fairness metrics, accounting for computational constraints and scalability requirements in real-world applications? (3) How can theories from information management guide the design of fairness metrics that reconcile conflicting stakeholder definitions of fairness, ensuring equitable treatment across diverse demographic groups while maintaining practical feasibility? Addressing these questions will facilitate the creation of more precise fairness metrics, ultimately leading to fairer and more inclusive AI systems across various applications.

*Advanced Debiasing Techniques:* Debiasing LLMs is critical, as post-deployment corrections are often costly, risky, and less effective (Gallegos et al., 2024). Although various debiasing techniques exist, their effectiveness can degrade significantly with minor modifications to datasets or evaluation settings (Sun et al., 2024). While these techniques represent a range of approaches, they may inadvertently increase bias over time (Tokpo et al., 2023). A further complexity arises from the tension between bias mitigation and algorithmic innovation. Prioritizing debiasing can discourage experimentation with novel designs, as developers may gravitate toward "safer" models with established methods, potentially stifling innovation.

To address these challenges, information management researchers, particularly those focusing on technical and quantitative approaches, should develop more robust and reliable debiasing methods. We propose the following research questions: (1) In debiasing finetuning, how can researchers balance the trade-off between over-tuning for accuracy and under-tuning for generalization? (2) What strategies can enable real-time debiasing in LLMs to adapt to evolving datasets while mitigating the risk of bias drift and performance degradation? (3) How can researchers identify and prioritize key features or characteristics that influence bias, ensuring effective mitigation without compromising scalability and model accuracy? Addressing these questions will pave the way for more effective debiasing strategies, ensuring LLM-driven applications sustain accuracy and fairness in real-world environments over time.

*Explainable AI for Transparency and Accountability*: Explainable AI (XAI) plays a vital role in enhancing the transparency and accountability of LLMs, making their decision-making processes more understandable and trustworthy to diverse users (Lawton, 2024). However, a core challenge in XAI design lies in balancing technical depth with user accessibility. Detailed explanations effectively serve technical stakeholders, such as developers and regulators, by providing the information needed for auditing and debugging. However, they can overwhelm or confuse non-technical stakeholders, such as end-users or decision-makers. Conversely, simplified explanations improve usability but risk omitting critical details and reducing information reliability and transparency.

To overcome these challenges, future research must focus on XAI techniques that balance these needs, offering actionable insights for technical users while remaining intuitive for non-experts. Additionally, information management scholars should prioritize seamlessly integrating XAI methods into existing information systems to enhance user trust without compromising system performance, scalability, or operational efficiency in demanding environments.

The following research questions explore the integration of Explainable AI in LLMs to enhance transparency and user trust: (1) How can XAI methods be integrated into LLMs to improve transparency while maintaining model performance? (2) How can XAI techniques be tailored for domain-specific applications (e.g., healthcare, finance, education) to ensure context-relevant transparency? Answering these questions will advance the deployment of Explainable AI in LLM-based decision systems, enhancing accountability, trust, and accessibility for both technical and non-technical users.

## 3.3 Policymaking and Social Impact
*Policy Development and Ethical Guidelines:* Establishing robust policies and ethical guidelines is crucial for ensuring the transparency, accountability, and fairness of LLMs. However, designing such frameworks is challenging due to the tension between universal standards and the need for context-specific flexibility. While standardized policies provide consistency and clarity, they may fail to address regional, cultural, or sector-specific nuances. Adding to this complexity is the rapid evolution of LLM technologies, which often outpaces existing policies and ethical frameworks, creating gaps that demand urgent attention. Information management scholars play a pivotal role in addressing these challenges by analyzing current policies, identifying gaps, and collaborating with policymakers to create practical and enforceable guidelines.

Key research questions on developing regulatory frameworks and ethical guidelines include: (1) How can information management researchers balance the tension between fostering innovation in LLMs and mitigating unintended societal consequences, particularly when ethical imperatives conflict with commercial pressures? (2) What ethical guidelines can direct the responsible integration of LLMs in high-stakes sectors like healthcare and finance, where conflicting demands for rapid deployment, accuracy, and inclusivity create unique regulatory approaches? Delving into these questions will help create policies that safeguard against bias, promote innovation, ensure responsible LLM deployment at scale, and support broader societal goals.

*Social Impact:* Evaluating the effectiveness of technical and policy interventions to reduce LLM bias necessitates a thorough assessment of their impact across various domains, including business, community, and society. A central challenge lies in balancing short-term organizational objectives—such as improving operational efficiency and profitability—with long-term societal goals of equity and inclusivity. For instance, interventions that prioritize fairness may require significant investments in time and resources, potentially delaying innovations or reducing immediate returns. By examining these tensions and their broader social implications, researchers can develop strategies that align organizational priorities with equitable outcomes across stakeholders. This approach ensures interventions genuinely promote inclusivity and fairness rather than inadvertently perpetuate disparities.

The following research questions focus on evaluating the societal impacts of LLM biases, particularly regarding marginalized communities and the trade-off between efficiency and equity: (1) How does bias propagate from foundational LLMs to downstream business applications, and what tensions arise when addressing these biases in business contexts prioritizing speed and scalability over fairness? (2) What trade-offs between efficiency and equity should be considered when evaluating the societal impact of LLM bias mitigation efforts, particularly in scenarios where mitigating bias may reduce profitability or create competitive disadvantages for organizations? Exploring these questions will inform strategies to foster inclusivity while balancing societal and organizational priorities, guiding the development of LLMs that prioritize efficiency and equity.

## 4. Applied Areas of Information Management Practice

LLMs have extensive applications in business practices, including human resources, healthcare, finance, marketing, etc. However, biases in LLMs can undermine the effectiveness and fairness of these applications, posing significant challenges to information management. This section identifies key areas where these biases manifest and suggests avenues for future research. We encourage scholars to apply the strategies proposed in Section 3 to explore these specific areas.

*Human Resources Management:* LLMs in human resource (HR) management, such as resume screening and performance evaluations, have been documented to perpetuate societal biases, perpetuating gender and racial discrimination that limits career opportunities for women and minority groups (Armstrong et al., 2024; Lucas, 2024). At the same time, hiring managers' efficiency goals might conflict with job applicants' concerns regarding fair representation in AI-driven recruitment systems. Overall, these LLM biases underscore the need for transparency, rigorous testing, and explainable AI methodologies to mitigate discrimination while balancing business constraints, such as minimizing adverse impacts or inefficiencies in achieving business outcomes. Key questions to address include: (1) How can bias in LLM-driven recruitment systems be designed to minimize bias and prevent gender and racial discrimination in hiring and evaluations? (2) What role can explainable AI play in enhancing the transparency and fairness of LLM-driven HR processes? Investigating these questions will help reduce discrimination in HR practices and provide a foundation for integrating explainable AI into HR systems.

*Healthcare Information Systems (HIS):* Biases in LLM-driven HIS can exacerbate health disparities and underserved marginalized populations (Santurkar et al., 2023). To ensure equitable healthcare, researchers should focus on real-time bias detection, transparency, and adapted methods for healthcare contexts (Levy et al., 2024). Relevant research questions include: (1) How can LLMs be tailored for healthcare systems to reduce biases that lead to disparities in patient treatment quality, particularly for marginalized groups? (2) How can researchers distinguish between intended differentiation, such as customizing treatment plans for specific populations (e.g., African Americans), and unintended differentiation that may lead to inequitable outcomes? Addressing these questions will advance LLM-driven healthcare systems that prioritize fairness, detect biases in real time, and prevent disparities in care delivery for marginalized groups.

*Financial Information Systems:* LLMs are reshaping the financial sector in areas like credit scoring, fraud detection, and algorithmic trading. However, their reliance on historical data often perpetuates systemic biases, disproportionately disadvantage marginalized groups, and create barriers to financial inclusion (O'Neil, 2017. Addressing these biases presents significant challenges. Financial institutions, such as banks, may prioritize efficiency and profitability, often clashing with fairness-focused interventions like fair lending practices and societal expectations of financial inclusion, as these measures may reduce model accuracy or increase costs. The severe consequences include exacerbated inequalities, regulatory violations, reputational damage, and even market destabilization in algorithmic trading. Key questions to address include: (1) How can information management scholars design fairness-aware LLM-based credit scoring frameworks that balance regulatory fairness mandates, organizational profitability, and equitable outcomes access for marginalized groups? (2) What strategies can mitigate real-time biases in LLM-driven algorithmic trading to ensure market stability and inclusivity while reconciling the conflicting priorities of high-speed decision-making and ethical accountability? Overcoming tensions in the questions requires balancing ethical considerations with practical constraints to build fair and transparent financial systems, which can be facilitated by collaborations among regulators, developers, and researchers.

*Marketing Applications:* While LLMs have transformed marketing by enabling personalized content creation, customer engagement, and market segmentation, tensions exist between achieving hyper-personalization and ensuring equitable marketing practices. Biases in LLM-driven marketing can reinforce stereotypes, exclude demographics, and perpetuate discriminatory practices (Hecks, 2024). For instance, biased algorithms may underrepresent minority groups in targeted advertisements or propagate harmful stereotypes in promotional content (PhummaArin, 2024), which risks alienating marginalized groups, damaging brand reputation, and reducing customer trust. Marketers face a trade-off between optimizing efficiency and addressing fairness. Highly optimized systems may inadvertently prioritize efficiency and cost-effectiveness over inclusivity, harming underrepresented communities. Key research questions include: (1) What frameworks can information management scholars develop to design LLM-based marketing systems that proactively minimize biases in audience targeting and content recommendations, ensuring equitable representation across diverse demographic groups while maintaining scalability and effectiveness? (2) What actionable methodologies can be designed to detect, analyze, and mitigate unintended biases in LLM-driven marketing systems while effectively balancing the competing priorities of hyper-personalization, customer engagement, inclusivity, and ethical accountability? To navigate these conflicting objectives, information management scholars should develop fairness-aware frameworks and implement explainable techniques to address biased patterns in content delivery and audience targeting.

*Customer Relationship Management (CRM) Systems:* CRM systems are pivotal for businesses to boost sales and revenue, improve customer retention, facilitate data-driven decision-making, and achieve operational efficiency. Chatbots powered by LLMs have recently significantly transformed customer interactions by offering accessibility and reducing service costs (Sachdeva et al., 2024). However, these chatbots often encounter fairness challenges due to biases inherent in the large training datasets used (Xue et al., 2023). For example, many voice assistants (e.g., Siri, Alexa) are designed with female personas, reinforcing gender stereotypes, while male personas are utilized for traditionally male-dominated tasks. These biases can impact user trust and satisfaction and entrench existing stereotypes. Therefore, it is pertinent to examine: (1) How can LLM-based CRM systems be optimized to avoid gender and cultural stereotypes in customer interactions, particularly in voice assistants and chatbots? (2) What role can explainable AI play in ensuring that LLM-driven CRM systems maintain fairness and transparency in customer service practices? Addressing these questions will help enhance the fairness and transparency of LLM-driven CRM systems, ensuring more equitable and unbiased customer service practices.

*Sentiment Analysis:* Sentiment analysis algorithms are extensively utilized within information management literature (Wankhade et al., 2022). LLMs can process large volumes of textual data and apply sentiment analysis algorithms for applications like customer feedback analysis, significantly reducing the need for coding and human validation. However, LLMs' inherent biases can lead to skewed results, influencing business decisions in areas such as product development and customer service. For instance, it has been observed that ChatGPT assigns more positive sentiments to countries with higher Human Development Index (HDI) scores, revealing a bias in fair representation (Georgiou, 2024). Additionally, many LLM-based models struggle with interpreting nuances such as irony or sarcasm and may favor certain languages (Buscemi and Proverbio, 2024). Information management researchers should focus on improving LLMs to address bias, enhance transparency, and ensure fair and accurate sentiment analysis. Pertinent research questions include: (1) How can biases in sentiment analysis algorithms using LLMs be addressed to ensure fair and accurate customer feedback analysis across different regions and cultures? (2) To what extent can LLMs be trained to better interpret nuances like sarcasm and irony, minimizing linguistic biases? Addressing these questions will lead to more inclusive and precise sentiment analysis models, allowing LLMs to better capture linguistic nuances while minimizing regional and cultural biases in customer feedback systems.

*Information Retrieval and Annotation:* Information retrieval from vast corpora presents a complex challenge for information management researchers. LLMs facilitate document analysis, relevant information extraction, and schema population with minimal human intervention, enhancing cost-effectiveness. In addition, LLMs can efficiently generate human-like text and reduce annotation time (Gururangan et al., 2020). However, biases in these applications may lead to unfair or inaccurate information extraction, adversely affecting decision-making and the quality and diversity of accessible information (Geerligs, 2024) Researchers must develop strategies to detect and mitigate these biases, ensuring fair and accurate information retrieval while implementing comprehensive policies to govern LLM use. The following research questions examine strategies for enhancing the accuracy and fairness of LLMs in information retrieval and annotation, especially when handling diverse and complex data corpora: (1) What finetuning methods can be developed to ensure that LLM models provide accurate results in information retrieval and annotation? (2) How can bias in LLM-assisted data annotation be continuously detected and corrected to prevent skewed outcomes? Addressing these questions will ensure that LLMs can detect and correct biases in information retrieval and annotation tasks, leading to more reliable datasets for future AI model training.

*Recommendation Systems:* Information management researchers have long focused on developing and optimizing recommendation systems to enhance performance and understand consumer behavior (Roy and Dutta, 2022). As LLMs become more integrated into these systems, the potential for bias increases (Saffarizadeh et al., 2024). For example, ChatGPT-based recommendation algorithms in e-commerce platforms may prioritize recent and popular content, such as movies, over older or less common genres and media (Deldjoo, 2024), thereby reinforcing existing user preferences and limiting exposure to diverse options. This phenomenon can adversely impact sales and user satisfaction. Consequently, researchers and practitioners must develop and implement strategies or guardrails to ensure equitable and accurate recommendations. Key research questions include: (1) How can LLM-based recommendation systems be designed to minimize bias and ensure fair content exposure across different user demographics and preferences? (2) What metrics can be established to evaluate the fairness and diversity of LLM-driven recommendation systems, particularly in e-commerce? Addressing these questions will contribute to developing more equitable and diverse recommendation systems, offering balanced content exposure and fair treatment of users across different demographics on e-commerce and other platforms.

*Other information systems:* GenAI-driven decision support systems used in organizations can exhibit biases that affect managerial decisions. For instance, biased risk assessment models might influence project approvals, investments, or resource allocation decisions, leading to unfair and adverse outcomes. This raises the question: (1) How can bias in LLM-driven decision support systems be mitigated to ensure fair resource allocation? Similarly, biases in fraud detection systems can result in higher false positive rates for transactions from certain demographic groups, leading to discriminatory practices where individuals from specific backgrounds are unfairly targeted for fraud investigations. Hence, another critical question is: (2) What methods can reduce false positive rates in LLM-based fraud detection systems to avoid discriminatory practices against specific demographic groups? Information management scholars have extensive opportunities to address the bias in these information systems, thereby enhancing the fairness of business outcomes.

LLM-driven information systems hold immense potential but are susceptible to biases that can skew research outcomes, unfair managerial decisions, discriminatory practices, and imbalanced resource allocations. These risks emphasize the need for information management scholars to focus on mitigating biases to ensure equitable business outcomes in extensive applied areas. By applying the implementation strategies outlined in 3.1, researchers and practitioners can safeguard against these biases. It is essential to prioritize transparency, fairness, and ethical principles in developing and applying LLM technologies, ensuring they contribute positively and equitably to society.

## 5. Conclusion

GenAI technologies have transformative potential for enhancing productivity and economic value across various business sectors (Yee and Chui, 2023). However, the inherent biases within these models, particularly in LLMs, raise significant ethical concerns. Trained on extensive internet text data, LLMs often mirror societal biases, stereotypes, and cultural assumptions, thereby influencing decision-making, entrenching stereotypes, and perpetuating inequalities. The dynamic nature of cultural and social norms further complicates the integration of these values into GenAI models, demanding a nuanced understanding of diverse and dynamic perspectives (Triandis, 2018).

This study highlights the critical issue of bias in GenAI and its profound implications for information management. Drawing on current research on bias detection, measurement, and mitigation, we propose future research directions specifically tailored for information management scholars. These include strategies for practical implementation and application in business contexts, guiding key research questions. We recognize that stakeholders in the LLM ecosystem, including developers, companies, users, and policymakers, often have varied and sometimes conflicting objectives. Hence, our research questions incorporate the need to balance these tensions to develop solutions that ensure more equitable LLMs.

Further, with the increasing technical and societal complexity of LLMs, we emphasize the need for interdisciplinary efforts. By integrating insights from computer science, ethics, law, and social sciences, information management scholars can develop more robust and comprehensive research designs. This interdisciplinary approach not only enhances the rigor and relevance of information management research but also ensures that solutions to LLM bias are holistic and effective. As such, our study provides a valuable framework for information management scholars to navigate and address the multifaceted challenges posed by LLM bias, which ultimately contributes to the development of fairer and more accountable AI systems.

Given the central role of information management in organizational operations, professionals must recognize the challenges associated with bias issues when adopting GenAI. This article fills a gap in the literature by providing insights and stimulating further discussions. We call for collaboration between practitioners and researchers to develop best practices in tackling these challenges.

Engaging actively in interdisciplinary research and collaboration, conducting continuous monitoring and longitudinal studies, and advocating for robust ethical frameworks and regulatory policies will enable scholars to build AI systems that are powerful, efficient, fair, and just. Information management curricula should also emphasize bias issues in GenAI, preparing students to navigate responsible implementation by understanding data integrity and ethical principles in GenAI practices. Through these efforts, we can contribute to a future where AI technologies drive positive societal outcomes and foster equity.

While GenAI offers significant business and social benefits, its rapidly evolving capabilities require innovation, vigilance, and ethical commitment. This research note addresses challenges like bias concerns and ethical issues, emphasizing the need for careful design and robust governance. Advances in fairness metrics, explainable AI, and human-in-the-loop systems will be crucial in achieving these goals. By prioritizing fairness, transparency, and inclusivity, information management scholars can ensure that AI technologies not only enhance organizational efficiency and decision-making but also promote social justice and equity.